\documentclass[12pt]{article}
\usepackage{times}
\usepackage{epsfig}
\usepackage{axodraw}
\usepackage{amsmath}
\usepackage{amsfonts}

\def\beq{\begin{equation}}
\def\eeq{\end{equation}}
\def\bea{\begin{eqnarray}}
\def\eea{\end{eqnarray}}

\def\nn{\nonumber}
\def\Eq#1{Eq.~(\ref{#1})}

\def\td#1{\widetilde{\delta}\left(#1\right)}
\def\thd#1{\tilde{\theta}\left(#1\right)}

\begin{document}
%%%%%%%%%%%%%%%%%%%%%%%%%%%%%%%%%%%%%%%%%%%%%%%%%%%%%%%%%%%%%%%%%%%%%

\begin{titlepage}
\renewcommand{\thefootnote}{\fnsymbol{footnote}}
\begin{flushright}
     IFIC/10-17
     \end{flushright}
\par \vspace{10mm}
\begin{center}
{\LARGE \bf
A Tree--Loop Duality Relation at Two Loops\\[0.5em] and Beyond}
\end{center}
\par \vspace{2mm}
\begin{center}
{\bf 
Isabella Bierenbaum~$^{(a)}$\footnote{E-mail: isabella.bierenbaum@ific.uv.es}, 
Stefano Catani~$^{(b)}$\footnote{E-mail: stefano.catani@fi.infn.it},
Petros Draggiotis~$^{(a)}$\footnote{E-mail: petros.drangiotis@ific.uv.es} and
Germ\'an Rodrigo~$^{(a)}$\footnote{E-mail: german.rodrigo@ific.uv.es}
}
\vspace{5mm}

${}^{(a)}$ Instituto de F\'{\i}sica Corpuscular, 
Universitat de Val\`{e}ncia -- 
Consejo Superior de Investigaciones Cient\'{\i}ficas, 
Apartado de Correos 22085, E-46071 Valencia, Spain \\
\vspace*{2mm}
${}^{(b)}$ INFN, Sezione di Firenze and
Dipartimento di Fisica, Universit\`a di Firenze,\\
I-50019 Sesto Fiorentino,
Florence, Italy \\
\vspace*{2mm}

\end{center}

\par \vspace{2mm}
\begin{center} {\large \bf Abstract} \end{center}
\begin{quote}
The duality relation between one--loop integrals and phase--space 
integrals, developed in a previous work, is extended to higher--order loops. 
The duality relation is realized by a modification of 
the customary $+i0$ prescription of the Feynman propagators, which compensates 
for the absence of the multiple--cut contributions that appear in the Feynman tree 
theorem. We rederive the duality theorem at one--loop order in a form that is more 
suitable for its iterative extension to higher--loop orders. 
We explicitly show its application to two-- and three--loop
scalar master integrals, and we discuss the structure of 
the occurring cuts and the ensuing results in detail.
\end{quote}

\par \vspace{5mm}

\vspace*{\fill}
\begin{flushleft}
October 25, 2010
\end{flushleft}
\end{titlepage}

\setcounter{footnote}{0}
\renewcommand{\thefootnote}{\fnsymbol{footnote}}

\section{Introduction}

With the start of the LHC, the physics of elementary particles enters a new
era, opening a powerful window to discover the Higgs boson and to explore new
interactions beyond the Standard Model (SM) at the TeV energy scale.
Precision theoretical predictions for background and signal multi--particle
hard scattering processes, in the SM and beyond, are mandatory for the
phenomenological interpretation of experimental data, and thus to achieve a
successful exploitation of the LHC physics programme.

While leading--order (LO) predictions of multi--particle processes at hadron
colliders in perturbative Quantum Chromodynamics (pQCD) provide, in general, a
rather poor description of experimental data, next--to--leading order (NLO) is
the first order at which normalizations, and in some cases the shapes, of
cross sections can be considered reliable,~\cite{Binoth:2010ra}.
Next--to--next--to leading order (NNLO), besides improving the determination
of normalizations and shapes, is also generally accepted to provide the first
serious estimate of the theoretical uncertainty in pQCD. Despite the
relatively smaller coupling, electroweak (EW) radiative NLO corrections might
also be sizable at the LHC,~\cite{Denner:2009gj,Kuhn:2007cv}.

Computing higher--order corrections in Quantum Field Theories, in particular
in QCD or in the EW sector of the SM, is highly challenging and substantially
demanding as the complexity increases with the number of external particles,
and the order in perturbation theory at which the hard scattering process must
be calculated in order to match the experimental precision.  In the recent
years, important efforts have been devoted to developing efficient methods
able to boost forward the calculational capability both at the multi--leg and
the multi--loop frontier.  Today, $2\to 4$ processes at NLO, either from
Unitarity based
methods,~\cite{Berger:2009zg,Melnikov:2009wh,Bevilacqua:2009zn}, or from a
more traditional Feynman diagrammatic approach,~\cite{Bredenstein:2010rs}, are
affordable and are even becoming standardized.  There has also been a lot of
progress concerning NNLO calculations
\cite{Bolzoni:2010xr,Catani:2010en,Catani:2009sm,Anastasiou:2007mz}.

In Ref.~\cite{Catani:2008xa}, a duality relation between one--loop integrals
and phase--space integrals has been demonstrated.  The duality relation is
suitable to numerically compute,~\cite{tanju}, multi--leg one--loop cross
sections in perturbative field theories (local and unitary).  It has analogies
with the Feynman tree theorem (FTT),~\cite{Feynman:1963ax,F2}, but involves
only single cuts of the one--loop Feynman diagrams.  The duality theorem
requires to properly regularize propagators by a complex Lorentz--covariant
prescription, which is different from the customary $+i0$ prescription of the
Feynman propagators.  The main consequence of this new prescription is that
the multiple cuts appearing in the FTT are avoided.

The computation of cross sections at NLO (or NNLO) requires the separate
evaluation of real and virtual radiative corrections.  Real (virtual)
radiative corrections are given by multi--leg tree--level (loop) matrix
elements to be integrated over the multi--particle phase--space of the physical
process.  The loop--tree duality at one--loop presented in
Ref.~\cite{Catani:2008xa}, as well as other methods relating one--loop and
phase--space integrals \cite{Soperetal,Kilian:2009wy,Moretti:2008jj}, have the
attractive feature that they recast the virtual radiative corrections in a
form that closely parallels the contribution of the real radiative
corrections.  This close correspondence can help to directly combine real and
virtual contributions to NLO cross sections.  In this paper, we extend the
loop--tree duality theorem derived in Ref.~\cite{Catani:2008xa} to
higher--order loops, as a first attempt towards extending the duality method
to the computation of cross sections at NNLO or even higher orders.
Preliminary results were presented in Ref.~\cite{isabella}.

The outline of the paper is as follows: In Section~\ref{sec:one-loop}, we
reconsider the tree--loop duality theorem at one--loop. This involves the
definition of dual propagators in addition to the Feynman, advanced and
retarded propagators commonly known. Afterwards, we reformulate the duality
theorem in a way which is more appropriate for extending it to higher loop
orders. This is done by providing functions of propagators of complete sets
of momenta, corresponding to the internal lines of the diagram. As for their
single--momenta analogues, it is likewise possible to infer relations amongst
these functions. The proof of the main relation which is crucial for the
extension to higher loop orders, is given in the Appendix~\ref{app:relations}. 
In Section~\ref{sec:two-loop}, we then provide a duality theorem for the
two--loop master diagram with $N$ external legs by using the previously
defined relations and iteratively applying the duality theorem to the
occurring loops.  We also discuss some subtleties involved, as well as
the structure of the result and occurring cuts. Furthermore, we 
derive a two--loop representation of the Feynman tree theorem. 
Having set up the
basic method in this way, we will continue with the four basic master
topologies at three loops in Section~\ref{sec:more-loops}, and show that the
method is indeed extendible to even higher loop orders in an iterative
manner. We end this section with a brief comment on the extension 
of the duality theorem at the amplitude level. 
In Section~\ref{sec:conclusion}, we conclude and provide an
outlook. In the Appendix~\ref{app:sunrise}, we provide a simple 
example of a two--loop scalar integral calculated from its dual 
representation.

\section{Duality relation at one loop}
\label{sec:one-loop}

In this section, we provide the basic quantities, definitions and relations
used in the rest of the paper, and sketch the steps for the derivation of the
tree--loop duality theorem at one--loop, as presented in
Ref.~\cite{Catani:2008xa}.  We will also rederive this tree--loop duality
theorem in a form which is more suitable for its extension to higher orders
by introducing the main formulas used for the iterative application of the
duality.

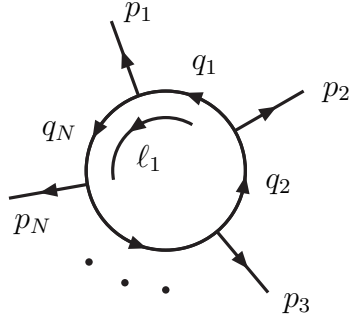
\begin{figure}[t]
\begin{center}
\vspace*{8mm}
\begin{picture}(120,110)(0,-10)
\SetWidth{1.2}
\BCirc(50,50){30}
\ArrowArc(50,50)(30,110,190)
\ArrowArc(50,50)(30,190,-50)
\ArrowArc(50,50)(30,-50,30)
\ArrowArc(50,50)(30,30,110)
\ArrowArc(50,50)(20,60,190)
\ArrowLine(39.74,78.19)(29.48,106.38)
\ArrowLine(75.98,65)(101.96,80)
\ArrowLine(69.28,27.01)(88.56,4.03)
\ArrowLine(20.45,44.79)(-9.09,39.58)
\Vertex(21.07,15.53){1.4}
\Vertex(34.60,7.71){1.4}
\Vertex(50,5){1.4}
\Text(44,55)[]{$\ell_1$}
\Text(40,110)[]{$p_1$}
\Text(65,90)[]{$q_1$}
\Text(115,80)[]{$p_2$}
\Text(93,45)[]{$q_2$}
\Text(10,65)[]{$q_N$}
\Text(0,30)[]{$p_N$}
\Text(100,0)[]{$p_3$}
\end{picture}
\end{center}
\vspace*{-6mm}
\caption{\label{f1loop} 
{\em Momentum configuration of the one--loop $N$--point scalar integral.}}
\end{figure}

Let us start by considering a general one--loop $N$--leg diagram, as
shown in Fig.~\ref{f1loop}, which is represented by the scalar integral:
%============================================================
\beq
\label{Ln}
L^{(1)}(p_1, p_2, \dots, p_N) =
\int_{\ell_1} \, \prod_{i=1}^{N} \, G_F(q_i)~.
\eeq
%============================================================
The four--momenta of the external legs are denoted $p_i$, $i \in
\{1,2,\ldots N\}$. All are taken as outgoing and ordered clockwise.  The
loop momentum is $\ell_1$, which flows anti--clockwise. The momenta of the
internal lines $q_i$, are defined as
%============================================================
\beq
\label{defqi}
q_i = \ell_1 + p_{1,i}~, \qquad i \in \alpha_1 = \{1,2,\ldots N\}~. 
\eeq 
%============================================================
As is commonly used, we define $p_{i,j}=p_i+p_{i+1}+\ldots+p_j$. Momentum
conservation is equivalent to $p_{1,N}=0$.  We use dimensionally regularized
integrals with the number of space--time dimensions equal to $d$, and
introduce the following shorthand notation:
%============================================================
\beq
\int_{\ell_i} \, \cdots  \equiv
- i \, \int \frac{d^d \ell_i}{(2\pi)^d} \, \cdots~.
\eeq
%============================================================
The space--time coordinates of any momentum $k_\mu$ are denoted as
$k_\mu=(k_0, {\bf k})$, where $k_0$ is the energy (time) component of
$k_\mu$. The Feynman propagators $G_F(q_i)$ in \Eq{Ln} have real internal
masses $m_i$:
%============================================================
\beq
G_F(q_i) = \frac{1}{q_i^2-m_i^2+i0}~.
\eeq
%============================================================
The derivation of the duality theorem is exactly the same regardless of the
internal lines being massive or massless ($m_i=0$), as long as the masses are
real. Non--vanishing internal real masses only account for a displacement of
the poles of the propagators along the real axis, which does not change the
derivation of the duality theorem, as will become obvious in the
following. Moreover, they do not alter the relationship between Feynman,
advanced, retarded and dual propagators, which is the basis of both, the
duality theorem as a duality to the FTT, as well as the extension of the
method to higher loop orders. The case of unstable particles with complex
masses has been discussed in detail in Ref.~\cite{Catani:2008xa}, and we do
not consider this possibility in the current paper.

%%============================================
%%============================================

\begin{figure}[t]
\begin{center}
\begin{picture}(350,110)(0,0)
\SetWidth{1.2}
\Line(0,50)(100,50)
\Line(200,50)(300,50)
\LongArrow(100,50)(150,50)
\LongArrow(300,50)(350,50)
\LongArrow(75,10)(75,90)
\LongArrow(275,10)(275,90)
\Text(10,85)[]{$G_F(q_i)$}
\Text(210,85)[]{$G_A(q_i)$}
\Text(140,10)[]{$q_{i,0}$ plane}
\Text(340,10)[]{$q_{i,0}$ plane}
\Text(110,35)[]{$\times$}
\Text(40,65)[]{$\times$}
\Text(315,65)[]{$\times$}
\Text(240,65)[]{$\times$}
\end{picture}
\end{center}
\caption{\label{fvsa}
{\em Location of the particle poles of the Feynman  (left)
and advanced (right) propagators $G_F(q_i)$ and $G_A(q_i)$
in the complex plane of the variable $q_{i,0}$.}}
\end{figure}
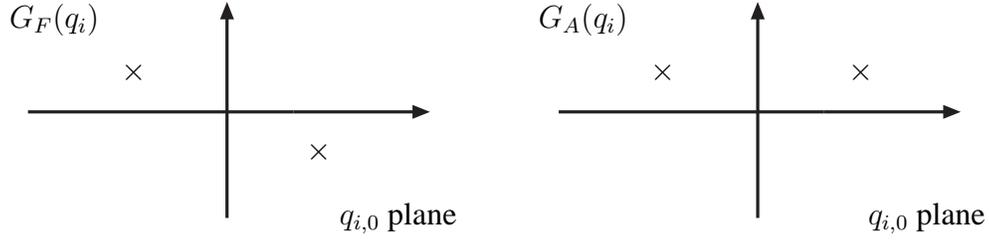

Besides the customary Feynman propagators $G_F(q_i)$, we also encounter
advanced, $G_A(q_i)$, and retarded, $G_R(q_i)$, propagators, defined by:
%============================================================
\beq 
G_A(q_i) = \frac{1}{q_i^2-m_i^2-i0\,q_{i,0}}~,\qquad
G_R(q_i) = \frac{1}{q_i^2-m_i^2+i0\,q_{i,0}}~.
\eeq
%============================================================
The Feynman, advanced, and retarded propagators only differ in
the position of the particle poles in the complex plane. Using $q_i^2 =
q_{i,0}^2 - {\bf q}_i^2$, we therefore find the poles of the Feynman and
advanced propagators in the complex plane of the variable $q_{i,0}$ at:
%============================================================
\beq
\label{fpole}
\left[ G_F(q_i)\right]^{-1} = 0  \:\:  \Longrightarrow \:\:
q_{i,0} = \pm  {\sqrt {{\bf q}_i^2 -m_i^2-i0}} \;\;
\:\:\: \mbox{and} \:\:\:
\left[ G_A(q_i)\right]^{-1} = 0 \:\: \Longrightarrow \:\:
q_{i,0} \simeq \pm  {\sqrt {{\bf q}_i^2 -m_i^2}} +i0~. 
\eeq
%============================================================
Thus, the pole with positive/negative energy of the Feynman propagator is
slightly displaced below/above the real axis, while both poles of the
advanced/retarded propagator, independently of the sign of the energy, are
slightly displaced above/below the real axis (cf. Fig.~\ref{fvsa}).  We
further define
%============================================================
\beq
\label{tdp}
\td{q_i} \equiv 2 \pi \, i \, \theta(q_{i,0}) \, \delta(q_i^2-m_i^2) 
= 2 \pi \, i \, \delta_+(q_i^2-m_i^2)~,
\eeq
%============================================================
where the subscript $+$ of $\delta_+$ refers to the on--shell mode with
positive definite energy, $q_{i,0}\geq 0$. Hence, the phase--space integral of
a physical particle with momentum $q_i$, i.e., an on--shell particle with
positive--definite energy, $q_i^2=m_i^2$, $q_{i,0}\geq 0$, reads:
%============================================================
\beq
\label{psm}
\int \frac{d^d q_i}{(2\pi)^{d-1}} \, \theta(q_{i,0}) \, \delta(q_i^2-m_i^2) 
\; \cdots \equiv \int_{q_i} \td{q_i} \; \cdots~.
\eeq

%%============================
%%============================

We continue by shortly recalling the duality theorem at one--loop order, which
was derived in Ref.~\cite{Catani:2008xa}. For a detailed discussion of all
definitions and steps, as well as subtleties related to them, we refer the
reader to this paper. In order to derive the duality theorem, one directly
applies the residue theorem to the computation of $L^{(1)}(p_1, p_2, \dots,
p_N)$ in \Eq{Ln}: Each of the Feynman propagators $G_F(q_i)$ has single poles
in both the upper and lower half--planes of the complex variable $\ell_{1,0}$.
Since the integrand is convergent when $\ell_{1,0}\to \infty$, by closing the
contour at $\infty$ in the lower half--plane and applying the Cauchy theorem,
the one--loop integral becomes the sum of $N$ contributions, each of them
obtained by evaluating the loop integral at the residues of the poles with
negative imaginary part belonging to the propagators $G_F(q_i)$.  The
calculation of the residue of $G_F(q_i)$ gives
%============================================================
\beq
\label{resGi}
{\rm Res}  [   G_F(q_i) ]_{{\rm Im}(q_{i,0}) < 0} 
= \int d \ell_{1,0} \, \delta_+(q_i^2-m_i^2)~,
\eeq
%============================================================
with $\delta_+(q_i^2-m_i^2)$ defined in Eq.~(\ref{tdp}).  This result shows
that considering the residue of the Feynman propagator of the internal line
with momentum $q_i$ is equivalent to cutting that line by including the
corresponding on--shell propagator $\delta_+(q_i^2-m_i^2)$.  The propagators
$G_F(q_j)$, with $j\neq i$, are not singular at the value of the pole of
$G_F(q_i)$ and can therefore be directly evaluated at this point, yielding to
%============================================================
\beq
\label{respre}
\prod_{j\neq i} \, G_F(q_j) 
\bigg|_{\substack{G_F(q_i)^{-1}=0\\ {\rm Im}(q_{i,0}) < 0}} = 
\prod_{j\neq i} \; G_D(q_i;q_j)~,
\eeq
%============================================================
where
%============================================================
\beq
G_D(q_i;q_j) = \frac{1}{q_j^2 -m_j^2 - i0 \,\eta (q_j-q_i)}~,
\eeq
%============================================================
is the so--called dual propagator, as defined in Ref.~\cite{Catani:2008xa},
with $\eta$ a {\em future--like} vector,
%============================================================
\beq
\label{etadef}
\eta_\mu = (\eta_0, {\bf \eta}) \;\;, \;\; \quad \eta_0 \geq 0, 
\; \eta^2 = \eta_\mu \eta^\mu \geq 0 \;\;,
\eeq
%============================================================
i.e.,~a $d$--dimensional vector that can be either light--like $(\eta^2=0)$ or
time--like $(\eta^2 > 0)$ with positive definite energy $\eta_0$.

Collecting the results from \Eq{resGi} and \Eq{respre}, the tree--loop duality
theorem at one--loop,~\cite{Catani:2008xa}, takes the final form
%============================================================
\bea
\label{oneloopduality}
L^{(1)}(p_1, p_2, \dots, p_N) 
&=& - \sum \, \int_{\ell_1} \; \td{q_i} \,
\prod_{\substack{j=1 \\ j\neq i}}^{N} \,G_D(q_i;q_j)~.
\eea 
%============================================================
Contrary to the FTT, \cite{Feynman:1963ax,F2}, \Eq{oneloopduality} contains
only single--cut integrals. Multiple--cut integrals, like those that appear in
the FTT, are absent thanks to modifying the original $+i0$ prescription of the
uncut Feynman propagators in \Eq{respre} by the new prescription $- i0 \,\eta
(q_j-q_i)$, which is named the `dual' $i0$ prescription or, briefly, the
$\eta$ prescription.  This is the main result of Ref.~\cite{Catani:2008xa}.
The dual $i0$ prescription arises from the fact that the original Feynman
propagator $G_F(q_j)$ is evaluated at the {\em complex} value of the loop
momentum $\ell_1$, which is determined by the location of the pole at
$q_i^2-m_i^2+i0 = 0$.  The $i0$ dependence of the pole of $G_F(q_i)$ modifies
the $i0$ dependence in the Feynman propagator $G_F(q_j)$, leading to the total
dependence as given by the dual $i0$ prescription. The presence of the vector
$\eta_\mu$ is a consequence of using the residue theorem and the fact that the
residues at each of the poles are not Lorentz--invariant quantities.  The
Lorentz--invariance of the loop integral is recovered after summing over all
the residues. Furthermore, in the one--loop case, the momentum difference
$\eta(q_j-q_i)$ is independent of the integration momentum $\ell_1$, and only
depends on the momenta of the external legs (cf. Eq.~(\ref{defqi})).

We now rederive the one--loop duality theorem \Eq{oneloopduality} by
exploiting the relationship between Feynman, advanced and dual propagators.
This will prove to be useful when extending the duality theorem to higher
orders. Using the elementary identity
%============================================================
\beq
\label{pvpid}
\frac{1}{x \pm i0} = {\rm PV}\left( \frac{1}{x} \right) \mp i \pi \, \delta(x) 
\;\;,
\eeq
%============================================================
where ${\rm PV}$ denotes the principal--value prescription, we can 
transform one kind of propagators into the other: 
%============================================================
\beq
\label{gavsgf}
G_{A}(q_i) = G_{F}(q_i)+\td{q_i} \;, \qquad
G_{R}(q_i) = G_{F}(q_i)+\td{-q_i}\;, \qquad
G_A(-q_i)  = G_R(q_i)~. 
\eeq
%============================================================
Dual and Feynman
propagators are related through,~\cite{Catani:2008xa},
%============================================================
\beq
\td{q_i} \; G_D(q_i;q_j) =
\td{q_i} \; \Bigl[ G_F(q_j) + \thd{q_j-q_i} \; \td{q_j} \Bigr]~,
\label{dovsfp}
\eeq 
%============================================================
with $\thd{q}=\theta(\eta q)$. 

In the following, we extend the definition of propagators 
of single momenta to combinations of propagators of
sets of internal momenta: Let $\alpha_k$ be any set of internal momenta with
$q_i,q_j \in \alpha_k$.  We then define Feynman, advanced, retarded and dual
propagator functions of this set $\alpha_k$ in the following way:
%============================================================
\beq
\label{eq:multi}
G_{F(A,R)} ( \alpha_k) = \prod_{i \in \alpha_k} G_{F(A,R)}( q_i)~, \qquad
G_D( \alpha_k) = \sum_{i \in \alpha_k} \, \td{ q_{i}} \, 
\prod_{\substack{j \in \alpha_k \\ j \neq i }} \, G_D( q_i; q_j)~.
\eeq
%============================================================
By definition, $G_D(\alpha_k)=\td{q_i}$, when $\alpha_k = \{i\}$ and thus
consists of a single four momentum. At one--loop order, $\alpha_k$ is
naturally given by all internal momenta of the diagram which depend on the
single integration loop momentum $\ell_1$, $\alpha_k=\{1,2,\ldots, N\}$.
However, let us stress that $\alpha_k$ can in principle be any set of
internal momenta. At higher order loops, e.g., several integration loop momenta
are needed, and we can define several loop lines $\alpha_k$ to label all the
internal momenta (cf. \Eq{lines}) where \Eq{eq:multi} will be used for these
loop lines or unifications of these. To simplify the notation, we also introduce
%============================================================
\beq
\label{eq:multiminus}
G_D(-\alpha_k) = \sum_{i \in \alpha_k} \, \td{-q_{i}} \, 
\prod_{\substack{j \in \alpha_k \\ j \neq i }} \, G_D(-q_i;-q_j)~, \qquad
G_{A} (-\alpha_k) = \prod_{i \in \alpha_k} G_{A}(-q_i) = G_{R} (\alpha_k)~,
\eeq
%============================================================
where the sign in front of $\alpha_k$ indicates that we have reversed the
momentum flow of all the internal lines in $\alpha_k$. For Feynman
propagators, moreover, $G_F(-\alpha_k)=G_F(\alpha_k)$.

In analogy to \Eq{gavsgf}, the following relation holds for any set of
internal momenta $\alpha_k$:
%============================================================
\beq
G_A(\alpha_k) = G_F(\alpha_k) + G_D(\alpha_k)~.
\label{eq:relevant}
\eeq
%============================================================
This is a non--trivial relation, considering \Eq{dovsfp}. Note that individual
terms in $G_D(\alpha_k)$ depend on the dual vector $\eta$, but the sum over
all terms contributing to $G_D(\alpha_k)$ is independent of it.  We leave the
detailed proof by induction of \Eq{eq:relevant} for the Appendix
\ref{app:relations}. \Eq{eq:relevant} is our main result for a straightforward
derivation of the duality theorem. 

Another crucial relation for the following is given by a formula that
allows to express the dual function of a set of momenta in terms of chosen
subsets. Consider the following set $\beta_N \equiv \alpha_1 \cup ... \cup
\alpha_N$, where $\beta_N$ is the unification of various subsets
$\alpha_i$. Solving for the dual part, \Eq{eq:relevant} then has the following
form:
%============================================================
\beq
\label{GAinGDGeneral}
G_D(\alpha_1 \cup \alpha_2 \cup ... \cup \alpha_N) 
= G_A(\alpha_1 \cup \alpha_2 \cup ... \cup \alpha_N) - G_F(\alpha_1 \cup
\alpha_2 \cup ... \cup \alpha_N)~.
\eeq
%============================================================
We continue by using the multiplicativity of $G_A(\beta_N)$ and $G_F(\beta_N)$,
as defined in \Eq{eq:multi}, to obtain:
%============================================================
\bea
\label{eq:GAinGDGeneralN}
G_D(\alpha_1 \cup \alpha_2 \cup ... \cup \alpha_N) 
&=& \prod_{i=1}^N G_A(\alpha_i) - \prod_{i=1}^N G_F(\alpha_i)\nn \\ 
&=& \prod_{i=1}^N \left[G_F(\alpha_i)+G_D(\alpha_i)\right] - \prod_{i=1}^N
G_F(\alpha_i)\nn \\ 
&=& \sum_{\substack{\beta_N^{(1)} \cup     \beta_N^{(2)}  = \beta_N}} \,
\prod_{\substack{i_1\in \beta_N^{(1)}}} \, G_D(\alpha_{i_1}) \,
\prod_{\substack{i_2\in \beta_N^{(2)}}} \,G_F(\alpha_{i_2})\,.  
\eea 
%============================================================
The sum runs over all partitions of $\beta_N$ into exactly two blocks
$\beta_N^{(1)}$ and $\beta_N^{(2)}$ with elements $\alpha_i,\linebreak i\in
\{1,...,N\}$, where, contrary to the usual definition, we include the case:
$\beta_N^{(1)} \equiv \beta_N$, $\beta_N^{(2)} \equiv \emptyset$. This
relation will be extensively used in the subsequent calculations. 
For the case of $N=2$, e.g.,
where $\beta_2 \equiv \alpha_1 \cup \alpha_2$, we have:
%============================================================
\beq
\label{eq:twoGD}
G_D(\alpha_1 \cup \alpha_2) 
= G_D(\alpha_1) \, G_D(\alpha_2)
+ G_D(\alpha_1) \, G_F(\alpha_2)
+ G_F(\alpha_1) \, G_D(\alpha_2)~.  
\eeq
%============================================================
Since in general relation (\ref{eq:GAinGDGeneralN}) holds for any
constellation of basic elements $\alpha_i$ which are sets of internal momenta,
one can look at these expressions in different ways, depending on the given
sets and subsets considered. If we define, for example, the basic subsets
$\alpha_i$ to be given by single momenta $q_i$, and since in that case
$G_D(q_i) = \td{q_i}$, Eq.~(\ref{eq:GAinGDGeneralN}) then denotes a sum over
all possible differing m--tuple cuts for the momenta in the set $\beta_N$,
while the uncut propagators are Feynman propagators. These cuts start from
single cuts up to the maximal number of cuts given by the term where all the
propagators of the considered set are cut.

Let us now return to the one--loop integral: Since advanced propagators have
poles with positive imaginary part only, we do not enclose any singularity by
closing the integration contour at $\infty$ in the lower half--plane, and
therefore a loop integral over advanced propagators vanishes. Thus, by using
\Eq{eq:relevant}, we find
%============================================================
\beq
0 = \int_{\ell_1} \, G_A(\alpha_1) = 
\int_{\ell_1} \, \left[ G_F(\alpha_1) + G_D(\alpha_1) \right]~,  
\label{eq:zero}
\eeq
%============================================================
where $\alpha_1$ as in \Eq{defqi} labels {\it all} internal momenta $q_i$. The
first term on the right--hand side of \Eq{eq:zero}, containing only Feynman
propagators, is the original one--loop integral. Therefore,
%============================================================
\beq
\label{eq:simpledual}
L^{(1)}(p_1, p_2, \dots, p_N) = - \int_{\ell_1} G_D(\alpha_1)~. 
\eeq
%============================================================
In this way, we directly obtain the duality relation between one--loop
integrals and single--cut phase--space integrals and hence \Eq{eq:simpledual}
can be interpreted as the application of the duality theorem to the given set
of momenta $\alpha_1$. It obviously agrees with
\Eq{oneloopduality}. Furthermore, by using \Eq{eq:GAinGDGeneralN} in its
refined form where the subsets $\alpha_i$ are given by the single momenta
$q_i$ of the inner lines of the one--loop integral, we rederive the FTT at
one--loop, namely the one--loop integral written in terms of multiple--cut
contributions and Feynman propagators:
%============================================================
\bea
\label{eq:FTT1loop}
L^{(1)}(p_1, p_2, \dots, p_N) 
&=& - 
\sum_{\alpha_1^{(1)}  \cup \alpha_1^{(2)} = \alpha_1} \, 
\int_{\ell_1} \, 
\prod_{\substack{i_1\in \alpha_1^{(1)}}} \, \td{q_{i_1}} \, 
\prod_{\substack{i_2\in \alpha_1^{(2)}}} \, G_F(q_{i_2})~.
\eea
%============================================================
The sum runs over all partitions of $\alpha_1$ as defined in
\Eq{eq:GAinGDGeneralN}, excluding the possibility to have a term with only
Feynman propagators.  The $m$--cut integral of the FTT is given by the sum of
the contributions from all partitions of $\alpha_1$, with $\alpha_1^{(1)}$
containing precisely $m$ elements\footnote{If the number of space--time
dimensions is $d$, then $m$ is limited to be $m \leq d$; the terms with
larger values of $m$ vanish, since the corresponding number of delta
functions in the integrand is larger than the number of integration
variables.}. 

The extension of the duality theorem and of the FTT from scalar loop integrals 
to full scattering amplitudes in the case of unitary, local field theories 
and in the occurrence of real masses is straightforward and has been discussed 
in detail in Ref.~\cite{Catani:2008xa}.

\section{Duality relation at two loops}
\label{sec:two-loop}

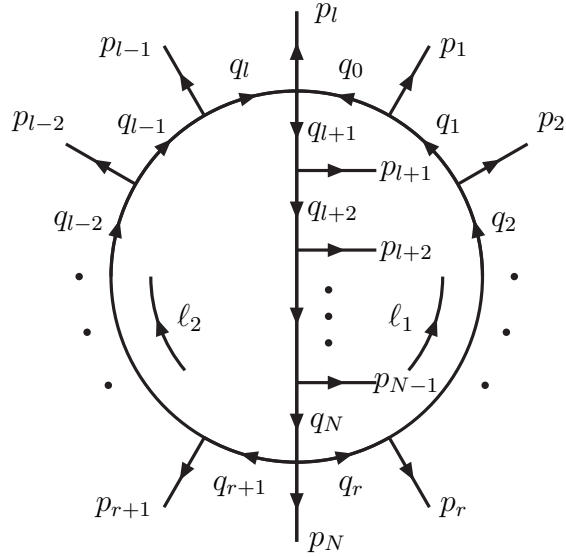
\begin{figure}[t]
\begin{center}

\begin{picture}(200,200)(0,-10)
\SetWidth{1.2}

\BCirc(100,100){70}
\Line(100,200)(100,0)
\ArrowLine(100,170)(100,200)
\ArrowLine(100,30)(100,0)

% loop momenta 
\ArrowArcn(100,100)(55,220,180)
\ArrowArc(100,100)(55,-40,0)

% middle line
\ArrowLine(100,60)(130,60)
\ArrowLine(100,140)(130,140)
\ArrowLine(100,110)(130,110)
\Vertex(112,95){1.4}
\Vertex(112,85){1.4}
\Vertex(112,75){1.4}

% right circle
\ArrowLine(135,161)(150,187)
\ArrowLine(161,135)(187,150)
\ArrowLine(135,39)(150,13)
\Vertex(179,79){1.4}
\Vertex(171,59){1.4}
\Vertex(182,100){1.4}

% left circle
\ArrowLine(65,161)(50,187)
\ArrowLine(39,135)(13,150)
\ArrowLine(65,39)(50,13)
\Vertex(21,79){1.4}
\Vertex(29,59){1.4}
\Vertex(18,100){1.4}

\Text(160,187)[]{$p_1$}
\Text(197,159)[]{$p_2$}
\Text(160,13)[]{$p_r$}
\Text(112,0)[]{$p_N$}
\Text(35,13)[]{$p_{r+1}$}
\Text(112,200)[]{$p_{l}$}
\Text(142,140)[]{$p_{l+1}$}
\Text(142,110)[]{$p_{l+2}$}
\Text(142,60)[]{$p_{N-1}$}
\Text(36,187)[]{$p_{l-1}$}
\Text(3,159)[]{$p_{l-2}$}

\ArrowArc(100,100)(70,60,90)
\ArrowArc(100,100)(70,30,60)
\ArrowArc(100,100)(70,0,30)
\ArrowArc(100,100)(70,-90,-60)

\ArrowArcn(100,100)(70,120,90)
\ArrowArcn(100,100)(70,150,120)
\ArrowArcn(100,100)(70,180,150)
\ArrowArcn(100,100)(70,270,240)

\ArrowLine(100,170)(100,140)
\ArrowLine(100,140)(100,110)
\ArrowLine(100,110)(100,60)
\ArrowLine(100,60)(100,30)

\Text(121,179)[]{$q_0$}
\Text(158,158)[]{$q_1$}
\Text(179,121)[]{$q_2$}
\Text(121,21)[]{$q_r$}
\Text(79,179)[]{$q_l$}
\Text(42,158)[]{$q_{l-1}$}
\Text(18,121)[]{$q_{l-2}$}
\Text(79,21)[]{$q_{r+1}$}
\Text(114,155)[]{$q_{l+1}$}
\Text(114,125)[]{$q_{l+2}$}
\Text(112,45)[]{$q_N$}
\Text(140,85)[]{$\ell_1$}
\Text(60,85)[]{$\ell_2$}

\end{picture}
\end{center}
\caption{\label{f2loop}
{\em Momentum configuration of the two--loop $N$--point scalar integral.}}
\end{figure}
%%============================================

We now turn to the general two--loop master diagram, as presented in
Fig.~\ref{f2loop}.  Again, all external momenta $p_i$ are taken as outgoing,
and we have $p_{i,j}=p_i+p_{i+1}+\ldots +p_j$, with momentum conservation
$p_{1,N} = 0$. The label $i$ of the external momenta is defined modulo $N$,
i.e., $p_{N+i} \equiv p_{i}$. Note, however, that one or both of the external
momenta attached to the four--leg vertices might be absent: $p_{\ell}=0$
and/or $p_N=0$. In the two--loop case, unlike at the one--loop order, the number of
external momenta might differ from the number of internal momenta. The loop momenta
are $\ell_1$ and $\ell_2$, which flow anti--clockwise and clockwise
respectively.  The momenta of the internal lines are denoted by $q_i$ and are
explicitly given by
%============================================================
\beq
\label{defqi2l}
q_i = \left\{
\begin{tabular}{ll}
$\ell_1+p_{1,i}$ & , $i \in \alpha_1$ \\
$\ell_2+p_{i,l-1}$ & , $i \in \alpha_2$ \\
$\ell_1+\ell_2+p_{i,l-1}$ &  , $i\in \alpha_3$ ~,
\end{tabular}
\right.  \eeq 
%============================================================
where $\alpha_k$, with $k=1,2,3$, are defined as the set of lines, propagators
respectively, related to the momenta $q_i$, for the following ranges of $i$:
%============================================================
\beq
\label{lines}
\alpha_1\equiv \{0,1,...,r\}~,\qquad \alpha_2\equiv \{r+1,r+2,...,l\}~,
\qquad \alpha_3\equiv \{l+1,l+2,...,N\}~.
\eeq
%============================================================
In the following, we will use $\alpha_k$
for denoting a set of indices or the set of the corresponding internal momenta
synonymously. Furthermore, we will refer to these lines often simply as the
``loop lines''.

We shall now extend the duality theorem to the two--loop case, 
by applying \Eq{eq:zero} iteratively. We consider first, in 
the most general form, a set of several loop lines $\alpha_1$ to $\alpha_N$
depending on the same integration momentum $\ell_i$, and find
%============================================================
\beq
\label{eq:ApplyDual}
\int_{\ell_i} \; G_F(\alpha_1 \cup \alpha_2 \cup ... \cup \alpha_N) 
= 
- \int_{\ell_i} \; G_D(\alpha_1 \cup \alpha_2 \cup ... \cup \alpha_N)~,
\eeq
%============================================================
which states the application of the duality theorem, \Eq{eq:zero}, to the set
of loop lines belonging to the same loop. \Eq{eq:ApplyDual} is the
generalization of the duality theorem found at one--loop to a single loop of a
multi--loop diagram. Each subsequent application of the duality theorem to
another loop of the same diagram will introduce an extra single cut, and by
applying the duality theorem as many times as the number of loops, a given
multi--loop diagram will be opened to a tree--level diagram. The duality
theorem, \Eq{eq:ApplyDual}, however, applies only to Feynman propagators, and
a subset of the loop lines whose propagators are transformed into dual
propagators by the application of the duality theorem to the first loop might
also be part of the next loop (cf., e.g., the ``middle'' line belonging to
$\alpha_3$ in Fig.~\ref{f2loop}). The dual function of the unification of
several subsets can be expressed in terms of dual and Feynman functions of the
individual subsets by using \Eq{eq:GAinGDGeneralN} (or \Eq{eq:twoGD}) though,
and we will use these expressions to transform part of the dual propagators
into Feynman propagators, in order to apply the duality theorem to the second
loop.

We are now ready for extending the duality relation to two loops.  Our starting
point is the expression for the two--loop $N$--leg scalar integral
%============================================================
\beq
\label{eq:L2}
L^{(2)}(p_1, p_2, \dots, p_N) =  \int_{\ell_1} \, \int_{\ell_2} \,
G_F(\alpha_1 \cup \alpha_2 \cup \alpha_3)~,
\eeq
%============================================================
where the momenta of the internal lines are specified in \Eq{defqi2l} and
\Eq{lines}. As stated before, we will apply the duality theorem sequentially
to the two different loops associated with the integration momenta $\ell_1$
and $\ell_2$. Starting with the first loop related to $\ell_1$, and hence to
the loop lines $\alpha_1$ and $\alpha_3$, and using \Eq{eq:ApplyDual}, we
obtain:
%============================================================
\beq
L^{(2)}(p_1, p_2, \dots, p_N) 
=  - \int_{\ell_1} \, \int_{\ell_2} \,
G_D(\alpha_1 \cup \alpha_3) \, G_F(\alpha_2)~. 
\eeq
%============================================================
We then use \Eq{eq:twoGD} for $G_D(\alpha_1 \cup \alpha_3)$, leading to 
%============================================================
\beq
\label{eq:L2firstloop}
L^{(2)}(p_1, p_2, \dots, p_N) 
=  - \int_{\ell_1} \, \int_{\ell_2} \,
\left\{ G_D(\alpha_1) \, G_D(\alpha_3) + G_D(\alpha_1) \, G_F(\alpha_3)
+ G_F(\alpha_1) \, G_D(\alpha_3) \right\} \, G_F(\alpha_2)~. 
\eeq
%============================================================
The first term of the integrand on the right--hand side of \Eq{eq:L2firstloop}
is the product of two dual functions, and therefore already contains double
cuts.  We do not modify this term further.  The second and third terms of
\Eq{eq:L2firstloop} contain {\it one} $G_D(\alpha_i)$ and hence single cuts
only.  We thus apply the duality theorem again, i.e., we use \Eq{eq:ApplyDual}
for $\ell_2$ in order to generate one more cut.  A subtlety arises at this
point since due to our choice of momentum flow, $\alpha_1$ and $\alpha_2$,
appearing in the third term of \Eq{eq:L2firstloop}, flow in the opposite
sense. Hence, in order to apply the duality theorem to the second loop we have
to reverse the momentum flow of one of these two loop lines. We choose to
change the direction of $\alpha_1$, namely $q_i\to -q_i$ for $i \in
\alpha_1$. This change of momentum flow is denoted by a sign in front of 
$\alpha_1$. Thus, applying \Eq{eq:ApplyDual} to the last two terms of
\Eq{eq:L2firstloop}, leads to
%============================================================
\bea
\label{AdvDual}
&& \!\!\!\!\!\!\!\!\!\!
L^{(2)}(p_1, p_2, \dots, p_N)  \nn \\
&& =   \int_{\ell_1} \int_{\ell_2} \, \left\{
- G_D(\alpha_1) \, G_F(\alpha_2) \, G_D(\alpha_3) 
+ G_D(\alpha_1) \, G_D(\alpha_2\cup \alpha_3)
+ G_D(\alpha_3) \, G_D(-\alpha_1\cup \alpha_2) \right\}~. \nn \\
\eea 
%============================================================
This is the dual representation of the two--loop scalar integral as a function
of double--cut integrals only, since all the terms of the integrand in
\Eq{AdvDual} contain exactly two dual functions as defined in \Eq{eq:multi}.
The integrand in \Eq{AdvDual} can then be reinterpreted as the sum over
tree--level diagrams integrated over a two--body phase--space.

The integrand in \Eq{AdvDual}, however, contains several dual functions of two
different loop lines, and hence dual propagators whose dual $i0$ prescription
might still depend on the integration momenta. This is the case for dual
propagators $G_D(q_i;q_j)$ where each of the momenta $q_i$ and $q_j$ belong to
different loop lines. If both momenta belong to the same loop line the
dependence on the integration momenta in $\eta(q_j-q_i)$ obviously cancels,
and the complex dual prescription is determined by external momenta only. The
dual prescription $\eta(q_j-q_i)$ can thus, in some cases, change sign within
the integration volume, therefore moving up or down the position of the poles
in the complex plane. To avoid this, we should reexpress the dual
representation of the two--loop scalar integral in \Eq{AdvDual} in terms of
dual functions of single loop lines. This transformation was unnecessary at
one--loop because at the lowest order all the internal momenta depend on the
same integration loop momenta; in other words, there is only a single loop
line.

Inserting \Eq{eq:twoGD} in \Eq{AdvDual} and reordering some terms, we arrive
at the following representation of the two--loop scalar integral
%============================================================
\bea
\label{AdvDualstar}
&& \!\!\!\!\!\!\!\!\!\!
L^{(2)}(p_1, p_2, \dots, p_N)  \nn \\
&& =   \int_{\ell_1} \int_{\ell_2} \, \left\{
  G_D(\alpha_1)  \, G_D(\alpha_2) \, G_F(\alpha_3) 
+ G_D(-\alpha_1) \, G_F(\alpha_2) \, G_D(\alpha_3)
+ G^*(\alpha_1)  \, G_D(\alpha_2) \, G_D(\alpha_3) \right\}~, \nn \\
\eea 
%============================================================
where
%============================================================
\beq
\label{Gstar1}
G^*(\alpha_k) \equiv G_F(\alpha_k) + G_D(\alpha_k) + G_D(-\alpha_k)~.
\eeq
%============================================================
This is the second main result of this paper.  In \Eq{AdvDualstar}, the
$i0$ prescription of all the dual propagators depends on external momenta
only. Through \Eq{Gstar1}, however, \Eq{AdvDualstar} contains also triple
cuts, given by the contributions with three $G_D(\alpha_k)$. The triple cuts
are such that they split the two--loop diagram into two disconnected
tree--level diagrams. By definition, however, the triple cuts
are such that there is no more than one cut per loop line $\alpha_k$. Since
there is only one loop line at one--loop, it is also clear why we did not
generate disconnected graphs at this loop order.
For a higher number of loops, we expect to
find at least the same number of cuts as the number of loops, and topology
dependent disconnected tree diagrams built by cutting up to all the loop lines
$\alpha_k$.  We explore this possibility at three loops in the next section.

Note that using \Eq{eq:relevant}, $G^*(\alpha_k)$ can also be expressed as
%============================================================
\beq
\label{Gstar2}
G^*(\alpha_k) = G_A(\alpha_k) + G_R(\alpha_k) - G_F(\alpha_k)~,
\eeq
%============================================================
which contains no cuts, although the imaginary prescription of
the advanced and retarded propagators still depends on the 
integration loop momenta. 

Finally, let us remark that from \Eq{AdvDualstar} we can obtain the FTT
representation of the two--loop scalar integral. More precisely, we can write
each $G_D(\alpha_k)$, for the lines $k \in \{1,2,3\}$, in terms of Feynman
propagators and multiple cuts of their constituting momenta, by using
\Eq{eq:GAinGDGeneralN} with the basic subsets $\alpha_i$ given by 
single momenta. After regrouping some terms, we obtain
%============================================================
\bea
\label{twoloop:feynman2}
&& L^{(2)}(p_1,\cdots,p_N) = 
\sum_{\substack{\alpha_k^{(1)} \cup \alpha_k^{(2)} = \alpha_k \\ k \in \{1,2,3\}}} \, 
\int_{\ell_1} \int_{\ell_2} \, \bigg\{
G_F(\alpha_1) \,
\prod_{i_1 \in \alpha_2^{(1)}} \, \td{q_{i_1}} \,
\prod_{i_2 \in \alpha_3^{(1)}} \, \td{q_{i_2}} \,
\prod_{\substack{i_3 \in \alpha_2^{(2)} \cup \alpha_3^{(2)}}} 
\, G_F(q_{i_3}) \nn \\
&& \qquad +  \, G_F(\alpha_2) \,
\prod_{i_1 \in \alpha_1^{(1)}} \, \td{-q_{i_1}} \,
\prod_{i_2 \in \alpha_3^{(1)}} \, \td{q_{i_2}} \,
\prod_{\substack{i_3 \in \alpha_1^{(2)} \cup \alpha_3^{(2)}}} 
\,G_F(q_{i_3}) \nn \\
&& \qquad +  \, G_F(\alpha_3) \,
\prod_{i_1 \in \alpha_1^{(1)}} \, \td{q_{i_1}} \,
\prod_{i_2 \in \alpha_2^{(1)}} \, \td{q_{i_2}} \,
\prod_{\substack{i_3 \in \alpha_1^{(2)} \cup \alpha_2^{(2)}}} 
\,G_F(q_{i_3}) \nn \\
&&  \qquad + 
\bigg( \prod_{i_1 \in \alpha_1^{(1)}} \, \td{q_{i_1}} 
+ \prod_{i_1 \in \alpha_1^{(1)}} \, \td{-q_{i_1}} \bigg)
\prod_{i_2 \in \alpha_2^{(1)}} \, \td{q_{i_2}} \,
\prod_{i_3 \in \alpha_3^{(1)}} \, \td{q_{i_3}} \,
\prod_{\substack{i_4 \in \alpha_1^{(2)} \cup \alpha_2^{(2)}\cup \alpha_3^{(2)}}} 
\,G_F(q_{i_4}) \bigg\}~, \nn \\
\eea
%============================================================
where the sums always run over the partitions of the index sets as defined
and used in \Eq{eq:GAinGDGeneralN}. We see that it consists of at least
double cuts up to a multiple cut of all the internal momenta.

\section{Duality relation beyond two loops}
\label{sec:more-loops}

%%============================================
\begin{figure}[t] 
\begin{center}
%
%Basket
%
\vspace*{8mm}
\begin{picture}(110,110)(0,-10)
\SetWidth{1.4}
\ArrowArc(50,50)(45,90,270)
\ArrowArc(50,50)(45,270,90)
\ArrowArcn(77,50)(52,240,120)
\ArrowArcn(23,50)(52,60,-60)
\Text(15,50)[]{$\alpha_4$}
\Text(40,50)[]{$\alpha_3$}
\Text(60,50)[]{$\alpha_2$}
\Text(85,50)[]{$\alpha_1$}
\Text(50,-20)[]{$(a)$}
\end{picture}
\hspace{.3cm}
%
%Zigzag
%
\begin{picture}(110,110)(0,-10)
\SetWidth{1.4}
\ArrowArc(50,50)(45,120,270)
\ArrowArc(50,50)(45,270,60)
\ArrowArcn(50,50)(45,120,60)
\ArrowLine(50,5)(28,88)
\ArrowLine(72,88)(50,5)
\Text(50,85)[]{$\alpha_5$}
\Text(20,30)[]{$\alpha_4$}
\Text(25,60)[]{$\alpha_3$}
\Text(75,60)[]{$\alpha_2$}
\Text(80,30)[]{$\alpha_1$}
\Text(50,-20)[]{$(b)$}
\end{picture}
\hspace{.3cm}
%
%Ladder
%
\begin{picture}(110,110)(0,-10)
\SetWidth{1.4}
\ArrowArc(50,50)(45,120,240)
\ArrowArc(50,50)(45,300,60)
\ArrowArcn(50,50)(45,120,60)
\ArrowArcn(50,50)(45,300,240)
\ArrowLine(28,12)(28,88)
\ArrowLine(72,88)(72,12)
\Text(50,15)[]{$\alpha_6$}
\Text(50,85)[]{$\alpha_5$}
\Text(15,40)[]{$\alpha_4$}
\Text(38,57)[]{$\alpha_3$}
\Text(62,43)[]{$\alpha_2$}
\Text(85,40)[]{$\alpha_1$}
\Text(50,-20)[]{$(c)$}
\end{picture}
\hspace{.3cm}
%
%Mercedes Star
%
\begin{picture}(110,110)(0,-10)
\SetWidth{1.4}
\ArrowArcn(50,50)(45,210,90)
\ArrowArc(50,50)(45,330,90)
\ArrowArcn(50,50)(45,330,210)
\ArrowLine(50,95)(50,50)
\ArrowLine(50,50)(89,26)
\ArrowLine(11,26)(50,50)
\Text(10,90)[]{$\alpha_6$}
\Text(50,15)[]{$\alpha_4$}
\Text(90,90)[]{$\alpha_1$}
\Text(22,47)[]{$\alpha_5$}
\Text(78,47)[]{$\alpha_3$}
\Text(65,77)[]{$\alpha_2$}
\Text(50,-20)[]{$(d)$}
\end{picture}
\end{center}
\vspace*{2mm}
\caption{\label{f3loop} {\em Master topologies of three--loop scalar
    integrals. Each internal line $\alpha_k$ can be dressed with an arbitrary
    number of external lines, which are not shown here.  }}
\end{figure}
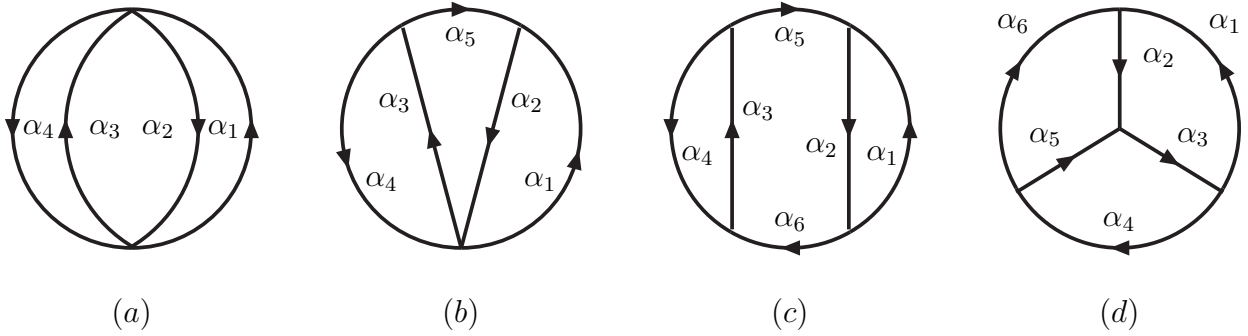

%============================================================
In this section, we will take a first look on the duality relation beyond the
two--loop order by considering the master topologies at three loops, as
represented in Fig.~\ref{f3loop}.  We obtain explicit representations of the
diagrams of Fig.~\ref{f3loop} by using the iterative method described in the
previous section.  Although the dual representations obtained in this section
are not unique, as the diagrams can be expressed in different ways in terms of
dual and Feynman propagators depending on the choice of lines whose momentum
flow is changed in the course of applying the duality theorem to each loop, we
have followed a systematic way in order to minimize the number of terms.
%============================================================

Diagrams \ref{f3loop}(a) - \ref{f3loop}(c) 
are in a certain sense of the same type, and will be
treated in the same way. We first cut these diagrams on the disjoint loops
assigned to the lines $\{\alpha_1,\alpha_2\}$ and
$\{\alpha_3,\alpha_4\}$. Considering the basket ball diagram (a) of
Fig.~\ref{f3loop}, for example, this means:
%============================================================
\bea
&& L^{(3)}_{\rm basket}(p_1, p_2, \dots, p_N) =  
\label{eq:basket1}
\int_{\ell_1} \int_{\ell_2} \int_{\ell_3} \,
G_D(\alpha_1 \cup \alpha_2)\; G_D(\alpha_3 \cup \alpha_4)~.
\eea
%============================================================
Remember that any dual function $G_D(\alpha_k)$ of any set of momenta, and hence any
application of the duality to a loop, contains at least one cut. Since we have
a product of two expressions of the dual type, all terms in the expansion of
this product in \Eq{eq:basket1} via \Eq{eq:twoGD} contain at least
two cuts. Terms with triple and more cuts are already in their final
form. They belong to the case where either all lines are dual, or one line is
of the type ``Feynman''. However, the double--cut terms, as, e.g.,
$G_D(\alpha_1) \, G_F(\alpha_2) \, G_F(\alpha_3) \, G_D(\alpha_4)$, stemming from the
combination of one Feynman propagator from each of the two different loops,
form a third loop, which still consists only of Feynman propagators and hence
still needs one more application of the duality theorem 
in order to generate the third cut:
%============================================================
\beq
\int_{\ell_1} \int_{\ell_2} \int_{\ell_3} \,
G_D(\alpha_1) \, G_F(\alpha_2) \, G_F(\alpha_3) \, G_D(\alpha_4)
\:\: \rightarrow \: \:
-\: \int_{\ell_1} \int_{\ell_2} \int_{\ell_3} \, 
G_D(\alpha_1) \, G_D(\alpha_2 \cup \alpha_3) \, G_D(\alpha_4)~.
\eeq
%============================================================
In the case of the zigzag diagram, \ref{f3loop}(b), 
or the ladder diagram, \ref{f3loop}(c), this third
loop consists of one more internal loop line: line $\alpha_5$ 
in the former case and the lines $\alpha_5$ and $\alpha_6$ in the latter. 
Hence the third loop now consists of exactly one loop line from the 
first and second loop, and these additional loop lines. Due to the
nature of the application of the duality through \Eq{eq:ApplyDual} and
\Eq{eq:GAinGDGeneralN}, we have to sum over all possibilities to build such
sets of loop lines fulfilling these properties. Additionally, we have to assure
that for each application of the duality the integration momentum runs in the
same sense and hence change the momentum--flow direction for some chosen
loop lines. Since each application of the duality generates one minus sign, we
obtain the following general result for diagrams \ref{f3loop}(a) - \ref{f3loop}(c):
%============================================================
\bea
\label{AtoC}
&& 
L^{(3)}_{\rm (a),(b),(c)}(p_1, p_2, \dots, p_N)  
=\int_{\ell_1} \int_{\ell_2} \int_{\ell_3} \,
G_D(\alpha_1 \cup \alpha_2)\; G_D(\alpha_3 \cup \alpha_4)\; G_F(\beta)\nn \\
&& =
\int_{\ell_1} \int_{\ell_2} \int_{\ell_3} \, \bigg\{  \bigg[
G_D(\alpha_2,\alpha_3,\alpha_4) \; G_F(\alpha_1) 
+G_D(\alpha_1,\alpha_3,\alpha_4)\; G_F(\alpha_2)
+G_D(\alpha_1,\alpha_2,\alpha_4)\; G_F(\alpha_3) \nn \\ && \qquad 
+G_D(\alpha_1,\alpha_2,\alpha_3)\; G_F(\alpha_4)
+G_D(\alpha_1,\alpha_2,\alpha_3,\alpha_4) \bigg]
G_F(\beta) 
\nn \\ && \qquad
-G_D(\alpha_1,\alpha_3)\; G_D(\alpha_2 \cup -\alpha_4 \cup \beta)
-G_D(\alpha_1,\alpha_4)\; G_D(\alpha_2 \cup \alpha_3 \cup \beta) \bigg.
\nn \\ && \qquad
-G_D(\alpha_2,\alpha_3)\; G_D(-\alpha_1 \cup -\alpha_4 \cup \beta)
-G_D(\alpha_2,\alpha_4)\; G_D(-\alpha_1 \cup \alpha_3 \cup \beta)
\bigg\}~,
\eea
%============================================================
where $\beta = \emptyset$ in the case of diagram \ref{f3loop}(a) (with
$G_F(\emptyset) = 1$), $\beta = \alpha_5$ for diagram \ref{f3loop}(b) and 
$\beta = \alpha_5 \cup \alpha_6$ in the case of the ladder diagram
\ref{f3loop}(c).
Note that for the ladder diagram, lines $\alpha_5$
and $\alpha_6$ depend on the same integration momentum and can therefore be
considered as a {\it single} loop line as defined in this context. Hence, in
this sense, the diagram naturally reduces to the zigzag--case, as long as
the relative sense of momentum flow in these lines is correct and stays
unchanged. 
For brevity, we defined the product of dual propagators as
$G_{D}(\alpha_1,...,\alpha_N)=\prod_{i=1}^N G_{D}(\alpha_i)$, in contrast to
$G_{D}(\alpha_1 \cup ... \cup \alpha_N)$, given in \Eq{eq:GAinGDGeneralN}. The
dual representation of the three--loop scalar integral in \Eq{AtoC}, contains
mostly triple cuts.  We have allowed for a single four--cut contribution in
order to make this expression more symmetric, but this term can be rewritten
in terms of triple--cut contributions with the help of \Eq{eq:twoGD}.

%============================================================
If we expand all existing dual functions in \Eq{AtoC} in terms of 
dual functions of single loop lines by using
\Eq{eq:GAinGDGeneralN}, we obtain, e.g., for diagram \ref{f3loop}(a):
%============================================================
\bea
\label{BasketExp}
&& L^{(3)}_{\rm basket}(p_1, p_2, \dots, p_N) 
= -\int_{\ell_1} \int_{\ell_2} \int_{\ell_3} \,
\bigg\{
G_D(\alpha_2,\alpha_3,-\alpha_4)\; G_F(\alpha_1)
+ G_D(\alpha_1,\alpha_3,-\alpha_4)\; G_F(\alpha_2) \nn \\ 
&&\qquad \bigg.
+ G_D(-\alpha_1,\alpha_2,\alpha_4)\; G_F(\alpha_3)
+ G_D(-\alpha_1,\alpha_2,\alpha_3)\; G_F(\alpha_4) \nn \\
&&\qquad \bigg.
+ G_D(-\alpha_1,\alpha_2,\alpha_3,\alpha_4)
+ G_D(\alpha_1,\alpha_2,\alpha_3,-\alpha_4)
+ G_D(-\alpha_1,\alpha_2,\alpha_3,-\alpha_4)
\bigg\}~.
\eea
%============================================================
In this expression, the complex dual prescription of all the dual propagators
depend on external momenta only, although at the price of generating
disconnected tree diagrams.  Similar results can be obtained for diagrams
\ref{f3loop}(b) and \ref{f3loop}(c).  There are up to four cuts for diagram
\ref{f3loop}(a) (cf. \Eq{BasketExp}), five cuts for diagrams \ref{f3loop}(b)
and six cuts for diagram \ref{f3loop}(c), although in this last case five cuts
are enough if $\alpha_5 \cup \alpha_6$ is considered as a single loop line.

Also the Mercedes star diagram \ref{f3loop}(d) 
can be expressed in terms of only three--cut contributions or in 
terms of three-- up to six-- cut contributions.
However, due to the non--planar nature of this diagram, the way
of obtaining its dual representation is slightly more involved, 
whereas the general idea as explained before stays the same. 
We achieve for the Mercedes star diagram the following 
dual representation:
%============================================================
\bea
\label{Mercedes}
&&\!\!\!\!\!\!\!\!
L^{(3)}_{\rm Mercedes}(p_1, p_2, \dots, p_N) \nn \\ &&=
\int_{\ell_1} \int_{\ell_2} \int_{\ell_3} \,
\bigg\{- G_D(\alpha_1,\alpha_2,\alpha_3) \, G_F(\alpha_4,\alpha_5,\alpha_6)
+ G_D(\alpha_3 \cup \alpha_4 \cup \alpha_5) \,
G_D(\alpha_1,\alpha_2) \, G_F(\alpha_6) \nn \\ && \bigg.
+ G_D(-\alpha_1 \cup \alpha_4 \cup \alpha_6) \,
G_D(\alpha_2,\alpha_3) \, G_F(\alpha_5) 
+ G_D(-\alpha_2 \cup \alpha_5 \cup -\alpha_6) \,
G_D(\alpha_1,\alpha_3) \, G_F(\alpha_4) \nn \\ && \bigg.
+ G_D(\alpha_1) [ 
G_D(\alpha_3 \cup \alpha_4) \, G_D(\alpha_5) \, G_F(\alpha_2\cup \alpha_6)
- G_D(\alpha_2 \cup \alpha_3 \cup \alpha_4 \cup \alpha_6) \, G_D(\alpha_5) 
\nn \\ && \bigg. \qquad
- G_D(\alpha_3 \cup \alpha_4) \, G_D(-\alpha_2 \cup \alpha_5 \cup -\alpha_6) ]
\nn \\ && \bigg. 
+ G_D(\alpha_2) [ 
G_D(-\alpha_1 \cup \alpha_6) \, G_D(\alpha_4) \, G_F(\alpha_3\cup \alpha_5)
- G_D(\alpha_1 \cup \alpha_3 \cup \alpha_5 \cup -\alpha_6) \, G_D(\alpha_4) 
\nn \\ && \bigg. \qquad
- G_D(-\alpha_1 \cup \alpha_6) \, G_D(\alpha_3 \cup \alpha_4 \cup \alpha_5) ]
\nn \\ && \bigg. 
+ G_D(\alpha_3) [ 
G_D(-\alpha_2 \cup \alpha_5) \, G_D(-\alpha_6) \, G_F(\alpha_1\cup \alpha_4)
- G_D(-\alpha_1 \cup -\alpha_2 \cup \alpha_4 \cup \alpha_5) \, G_D(-\alpha_6) 
\nn \\ && \bigg. \qquad
- G_D(-\alpha_2 \cup \alpha_5) \, G_D(-\alpha_1 \cup \alpha_4 \cup \alpha_6) ]
\bigg\}~.
\eea
%============================================================

From the inspection of the three--loop case and from the general derivation
of the method, it seems obvious how to extend it to even higher loop orders.
\Eq{eq:GAinGDGeneralN} can also be used to obtain the FTT representation of
scalar integrals or scattering amplitudes at any loop order.

The duality relation can be extended to evaluate not only
scalar loop integrals, as discussed so far,
but also complete Feynman diagrams.
The extension of the one-loop duality
relation from scalar integrals
to Feynman diagrams  was discussed in details
in Ref.~\cite{Catani:2008xa}. This extension relies on the
simple observation that the duality relation acts only on the Feynman
propagators of the loop,
leaving unchanged all the other factors in the Feynman diagram.
This is valid in any unitary and local field theories.
In spontaneously broken gauge theories, it holds
in the 't Hooft-Feynman gauge and in the unitary gauge.
In unbroken gauge theories, the duality
relation is valid in the 't Hooft-Feynman gauge, and in physical
gauges where the gauge vector $n^\nu$ is orthogonal to the
dual vector $\eta^\mu$, i.e., $n\cdot \eta=0$. This
excludes gauges where $n^\nu$ is time-like.
At one-loop order,
this choice
of gauges avoids the appearance of extra unphysical
gauge poles, which in other gauges (e.g. the time-like axial gauge)
are also poles of the second order.
Within the same choice of gauges, additional (unphysical) gauge poles are
absent also at higher-loop level, and the duality relation
can be straightforwardly extended from scalar integrals to Feynman
diagrams.
 
In Ref.~\cite{Catani:2008xa}, it was also shown how the one-loop duality
relation can directly be expressed at the level of full scattering
amplitudes
(or, more precisely, off-shell Green's functions).
The derivation of the duality between one-loop and tree-level
scattering amplitudes requires a detailed discussion of some issues
related
to tadpole and self-energy configurations. These (and related) issues 
become more delicate at higher-loop levels. We do not pursue further
on this point in this paper, and we postpone detailed investigations to
further studies.

%%%%%%%%%%%%%%%%%%%%%%%%%%%%%%%%%%%%%%%%%%%%%%%%%%%%%%%%%%%%%%%%%%%%
%%%%%%%%%%%%%%%%%%%%%%%%%%%%%%%%%%%%%%%%%%%%%%%%%%%%%%%%%%%%%%%%%%%%
%%%%%%%%%%%%%%%%%%%%%%%%%%%%%%%%%%%%%%%%%%%%%%%%%%%%%%%%%%%%%%%%%%%%
%%%%%%%%%%%%%%%%%%%%%%%%%%%%%%%%%%%%%%%%%%%%%%%%%%%%%%%%%%%%%%%%%%%%

\section{Conclusion and Outlook}
\label{sec:conclusion}

We have rederived the tree--loop duality theorem at one--loop order, which was
introduced in Ref.~\cite{Catani:2008xa}, in a way which is more suitable for
extending it to higher loop orders. By iteratively applying the duality
theorem, we have given explicit representations of the two-- and three--loop
scalar integrals. The method, however, is easily extendible to higher loop
orders beyond three loops. In general, the dual representation of the loop
integrals can be written as a sum of terms with exactly the same number of
cuts as the number of loops, and in such a way that the loop diagram is opened
to a tree--level diagram. However, this requires to deal with uncut
propagators with complex dual $i0$ prescription depending on the integration
momenta, and thus with complex dual prescriptions that might change sign
within the integration volume.  Dual representations of the loop integrals
with complex dual prescription depending only on the external momenta can be
obtained at the cost of introducing extra cuts, which break the loop integrals
into disconnected diagrams.  This is a new feature of the duality theorem
beyond one--loop, which does not happen at the lowest order. The number of
extra cuts to be taken into account depends on the topology of the loop
diagram. The maximal number of cuts agrees with the number of loop lines, and
the cuts are such that it does not appear more than a single cut for each
internal loop line. These general facts are true for the application of the
duality relation to diagrams with an arbitrary number of loops. The results
presented in this paper can also be used to obtain the FTT representation of
diagrams at higher orders.

The dual representations obtained in this paper are valid as far as only
single poles are present when the residue theorem is applied. At one--loop
order, the propagators of the gauge bosons might generate unphysical poles, or
even higher order poles. Those non--single poles can be avoided by a convenient
choice of the gauge or of the dual vector~\cite{Catani:2008xa}. At two-- or
higher loop orders, however, higher order poles might appear when diagrams
with selfenergy insertions (nested or disjoint) are considered. At two loops
this happens when two of the loop lines are made of single propagators, and no
external momenta are attached to any of the two four-leg vertices of
Fig.~\ref{f2loop}. At higher order loops, there are many more possible
topologies showing this feature. Extending the duality theorem to this kind of
diagrams requires to evaluate the contribution of the higher order poles,
which depends on the topology of the diagram and on the nature of the internal
propagators and on the form of the interaction vertices. 
Explicity studies of these loop diagrams are left to future investigations.

{\it Note added:} After completion of this paper, a
work~\cite{CaronHuot:2010zt} dealing with similar topics appeared.  The author
uses retarded boundary conditions, and obtains some combinatorial factors
weighting the different terms contributing to the loop-tree duality relation.
We haved checked explicitly, in the two-loop case, that such combinatorial
factors are the result of averaging over the different dual countertparts of
the same loop integral obtained by permuting the loop lines
$\alpha_i$. Although many of the terms obtained in this way are either
equivalent or can be related to each other after shifting the loop momenta, a
larger amount of terms than by using a single dual counterpart are needed to
be summed up for the same loop integral or Feynman diagram.  By using
\Eq{eq:extra1} or \Eq{eq:extra2} dual propagators can be expressed in terms of
advanced or retarded propagators in a straightforward way, leading to an
equivalent loop-tree duality relation as presented in
Ref.~\cite{CaronHuot:2010zt}.  In our paper we have not followed this
procedure, since a main feature of our duality relation is that the `$i0$'
prescription of the dual propagators depends only on the external momenta.

\section*{Acknowledgments} 

This work was supported by the Ministerio de Ciencia e Innovaci\'on under
Grant No. FPA2007-60323, by CPAN (Grant No. CSD2007-00042), by the Generalitat
Valenciana under Grant No. PROMETEO/2008/069, by the European Commission under
contracts FLAVIAnet (MRTN-CT-2006-035482) and HEPTOOLS (MRTN-CT-2006-035505),
and by INFN-MICINN agreement under Grants No. ACI2009-1061 and
FPA2008-03685-E.

\appendix

\section{Derivation of some algebraic relations}
\label{app:relations}

In this Appendix, we prove by induction several algebraic relations 
that have been used in the text. The basic ingredient of the proof 
is the following relation:
%============================================================
\beq
\theta(\lambda_1) \,\theta(\lambda_1+\lambda_2) \,\dots 
\,\theta(\lambda_1+\lambda_2+ \dots+\lambda_{n-1} ) + {\rm cyclic \;\; perms.}
 = 1~,
\label{eq:algebraic}
\eeq 
%============================================================
that holds for any set of $n$ real variables $\lambda_i$, 
with $i=1,2,\dots,n$, such that 
%============================================================
\beq
\sum_{i=1}^n \lambda_i = 0~.
\label{lacon}
\eeq
%============================================================
Relation (\ref{eq:algebraic}) was proven in Appendix B of 
Ref. \cite{Catani:2008xa}. It applies, in particular, to
$\lambda_i=\eta p_i$, and follows from momentum 
conservation $\sum_{i=1}^n p_i = 0$, where $p_i$ are 
external momenta. In the following, we will use \Eq{eq:algebraic}
by setting $\lambda_i=\eta(q_i-q_{i+1})$ for $i \in \{1,...,n\}$, 
with $(n+i) \equiv i\; \mbox{mod}\; n$.
The four--momenta $q_i$ are any arbitrary set of internal momenta, 
and the real variables $\lambda_i=\eta(q_i-q_{i+1})$ might still 
depend on the loop momenta $\ell_1$ and $\ell_2$. By construction, 
however, \Eq{lacon} is automatically fulfilled. Thus,  
\Eq{eq:algebraic} can also be written as:
%============================================================
\beq
\sum_{i=1}^{n}
\prod_{\substack{j=1 \\ j \neq i}}^n \thd{q_i-q_j} =  
\sum_{i=1}^{n} 
\prod_{\substack{j=1 \\ j \neq i}}^n
\thd{q_j-q_i} = 1~.
\label{eq:algebraic2}
\eeq
%============================================================

We start by deriving the following algebraic identity: 
%============================================================
\beq
G_D(\alpha_k) = G_A(\alpha_k) - G_F(\alpha_k)~,
\label{eq:GDGAGF}
\eeq
%============================================================
where
%============================================================
\beq
G_D(\alpha_k) = \sum_{i \in \alpha_k} \, \td{q_{i}} \, 
\prod_{\substack{j \in \alpha_k \\ j \neq i }} \, G_D(q_i;q_j)~, \qquad
G_{A(F)} (\alpha_k) = \prod_{i \in \alpha_k} G_{A(F)}(q_i)~.
\eeq
%============================================================
as defined in \Eq{eq:multi}. Remember that $G_D(\alpha_k)=\td{q_i}$ when
$\alpha_k$ consists of a single four momentum $q_i$.  Using the identity
$G_A(q_i)=G_F(q_i)+\td{q_i}$, the right-hand side of \Eq{eq:GDGAGF}, can also
be written as
%============================================================
\beq
G_A(\alpha_k) - G_F(\alpha_k) = 
\sum_{\alpha_k^{(1)}  \cup \alpha_k^{(2)} = \alpha_k} \, 
\prod_{\substack{i_1\in \alpha_k^{(1)}}} \, \td{q_{i_1}} \, 
\prod_{\substack{i_2\in \alpha_k^{(2)}}} \, G_F(q_{i_2})~, 
\label{eq:GAGF}
\eeq
%============================================================
where the sum runs over all possible partitions of $\alpha_k$ into exactly two
subsets $\alpha_k^{(1)}$ and $\alpha_k^{(2)}$. Additionally, we allow for the
subset $\alpha_k^{(2)}$ to be empty, but the subset $\alpha_k^{(1)}$ always
contains at least one element, this means that a term with only Feynman
propagators is excluded.  The first non trivial case for \Eq{eq:GDGAGF} occurs
for $\alpha_k=\{1,2\}$.  By using Eq.~(\ref{dovsfp}), and
$\thd{q_1-q_2}+\thd{q_2-q_1}=1$, it is straightforward to prove that
\beq
\td{q_1} \, G_D(q_1;q_2) + \td{q_2} \, G_D(q_2;q_1)
= \td{q_1} \, G_F(q_2) + \td{q_2} \, G_F(q_1) + \td{q_1} \, \td{q_2}~. 
\eeq 

Let us now assume that relation (\ref{eq:GDGAGF}) is correct for 
$N$ loop momenta, $\alpha_k=\{1,\ldots, N\}$, 
and show that it is valid for $\alpha_k^{N+1}=\alpha_k \cup \{N+1\}$. 
We replace the dual propagators $G_D(q_i;q_{N+1})$ 
appearing in $G_D(\alpha_k^{N+1})$ by using again \Eq{dovsfp}.
We obtain: 
%============================================================
\begin{eqnarray}
G_D(\alpha_k^{N+1}) &=& 
\sum_{i=1}^{N+1} \, \td{q_{i}} \, 
\prod_{\substack{j =1 \\ j \neq i }}^{N+1} \, G_D(q_i;q_j)  = 
G_D(\alpha_k) \, G_F(q_{N+1}) \nn \\
&+& \td{q_{N+1}} \bigg(
\sum_{i =1}^{N} \, \thd{q_{N+1}-q_i} \, \td{q_{i}} \, 
\prod_{\substack{j =1 \\ j \neq i }}^{N} \, G_D(q_i;q_{j})\, 
+ \prod_{\substack{j =1 \\ j \neq i}}^{N} \, G_D(q_{N+1};q_j) \bigg)~. 
\label{eq:IndStep2}
\end{eqnarray}
%============================================================
Assuming that \Eq{eq:GDGAGF} and \Eq{eq:GAGF}  
are valid for $N$ elements, 
the first term in the right-hand side of \Eq{eq:IndStep2}, 
which is proportional to $G_F(q_{N+1})$, becomes
%============================================================
\beq
G_D(\alpha_k) \, G_F(q_{N+1}) = 
\left[ G_A(\alpha_k) - G_F(\alpha_k) \right] G_F(q_{N+1})~. 
\label{eq:induction1}
\eeq
%============================================================
For the remaining terms in \Eq{eq:IndStep2}, 
which are proportional to $\td{q_{N+1}}$, we apply again 
\Eq{dovsfp} to all dual propagators. After some algebra, 
we find: 
%============================================================
\bea
&& \!\!\!\!\!\!\!\!\!\!\!
G_D(\alpha_k^{N+1}) - G_D(\alpha_k) \, G_F(q_{N+1}) = \nn \\ &&
 \td{q_{N+1}} \Bigg( G_F(\alpha_k)  
+ \sum_{\alpha_k^{(1)}  \cup \alpha_k^{(2)} = \alpha_k} \, 
\tilde \Theta(\alpha_k^{(1)}\cup\{N+1\}) \,
\prod_{\substack{j_1\in \alpha_k^{(1)}}} \, \td{q_{j_1}} \, 
\prod_{\substack{j_2\in \alpha_k^{(2)}}} \, G_F(q_{j_2}) \Bigg)~, 
\label{eq:induction2}
\eea
%============================================================
with
%============================================================
\beq
\tilde \Theta(\alpha_k^{(1)}\cup\{N+1\}) = 
\sum_{\substack{i_1\in \alpha_k^{(1)}\cup\{N+1\}}} \Bigg( 
 \prod_{\substack{i_2\in \alpha_k^{(1)}\cup\{N+1\} \\ i_2 \ne i_1}} 
\thd{q_{i_1}-q_{i_2}} \Bigg)~,
\eeq
%============================================================
where the sum runs over all possible products of $\tilde \theta$
functions that can be constructed with the four--momenta in the 
set $\alpha_k^{(1)}$ and $q_{N+1}$. Obviously, from 
%============================================================
\Eq{eq:algebraic2},
\beq
\tilde \Theta(\alpha_k^{(1)}\cup\{N+1\}) = 1~,
\label{eq:induction3}
\eeq
%============================================================
for all possible partitions of $\alpha_k$ into $\alpha_k^{(1)}$ and
$\alpha_k^{(2)}$. Collecting the results from \Eq{eq:induction1},
\Eq{eq:induction2}, and \Eq{eq:induction3}, we finally obtain
%============================================================
\bea
G_D(\alpha_k^{N+1}) &=& G_D(\alpha_k) \, G_F(q_{N+1}) +  
\td{q_{N+1}} \left( G_F(\alpha_k)  + G_D(\alpha_k) \right) \nn \\ 
&=& G_A(\alpha_k) \, G_A(q_{N+1}) -  G_F(\alpha_k) \, G_F(q_{N+1})~, 
\eea
%============================================================
as we wanted to demonstrate. 

Another useful relation of advanced and Feynman propagators 
is the following:
%============================================================
\beq
G_D(\alpha_k) = \sum_{i\in \alpha_k}
\Bigg( \prod_{\substack{j\in \alpha_k \\ j < i}} G_A(q_j) \Bigg) \,
\td{q_i} \,
\Bigg( \prod_{\substack{l\in \alpha_k \\ l > i}} G_F(q_l) \Bigg)~.
\label{eq:extra1}
\eeq
%============================================================
We do not attempt to present a detailed proof here, as \Eq{eq:extra1}
can be straightforwardly derived by reordering some terms from 
\Eq{eq:GAGF}. Analogously, we also have 
%============================================================
\beq
G_D(-\alpha_k) = G_R(\alpha_k) - G_F(\alpha_k)~,
\eeq
%============================================================
and
%============================================================
\beq
G_D(-\alpha_k) = \sum_{i\in \alpha_k}
\Bigg( \prod_{\substack{j\in \alpha_k \\ j < i}} G_R(q_j) \Bigg) \,
\td{-q_i} \,
\Bigg( \prod_{\substack{l\in \alpha_k \\ l > i}} G_F(q_l) \Bigg)~,
\label{eq:extra2}
\eeq
%============================================================
which also hold straightforward by changing the momentum flow of
the four--momenta from $\alpha_k$ in \Eq{eq:GDGAGF} and \Eq{eq:extra1}
and taking into account that $G_F(-q_i)=G_F(q_i)$ and 
$G_A(-q_i)=G_R(q_i)$.

%%%%%%%%%%%%%%%%%%%%%%%%%%%%%%%%%%%%%%%%%%%%%%%%%%%%%%%%
%%%%%%%%%%%%%%%%%%%%%%%%%%%%%%%%%%%%%%%%%%%%%%%%%%%%%%%%
%%%%%%%%%%%%%%%%%%%%%%%%%%%%%%%%%%%%%%%%%%%%%%%%%%%%%%%%
%%%%%%%%%%%%%%%%%%%%%%%%%%%%%%%%%%%%%%%%%%%%%%%%%%%%%%%%
\section{Massless sunrise two--loop two--point function} 
\label{app:sunrise}

%%============================================
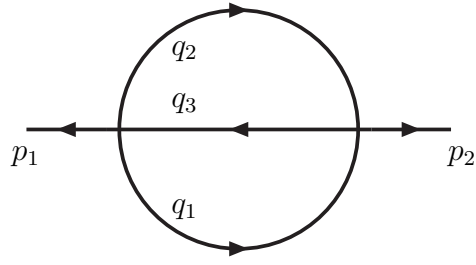
\begin{figure}[htb]
\begin{center}

\begin{picture}(200,100)(0,0)
\SetWidth{1.4}

\ArrowArc(100,50)(45,180,0)
\ArrowArcn(100,50)(45,180,0)
\ArrowLine(150,50)(50,50)
\ArrowLine(50,50)(20,50)
\ArrowLine(150,50)(180,50)

\Text(80,20)[]{$q_1$}
\Text(80,60)[]{$q_3$}
\Text(80,80)[]{$q_2$}
\Text(20,40)[]{$p_1$}
\Text(185,40)[]{$p_2$}

\end{picture}
\end{center}
\caption{\label{Sun2loop}
{\em Sunrise two--loop two--point function.
}}
\end{figure}
%%============================================

We consider, as explicit example of the application of the 
duality relation at two--loops, the massless sunrise two--loop two--point 
function (Fig.~\ref{Sun2loop}). From \Eq{AdvDualstar}, 
the dual representation of the sunrise two--loop scalar integral is
given by 
\beq
L^{(2)}(p_1,p_2) = \int_{\ell_1} \, \int_{\ell_2}
\left\{ \td{q_1} \, \td{q_2} \, G_F(q_3) + \td{-q_1} \, G_F(q_2) \, \td{q_3} 
+ G^*(q_1) \, \td{q_2} \, \td{q_3} \right\}~.
\eeq
After replacing $G^*(q_1)=G_F(q_1)+\td{q_1}+\td{-q_1}$, and 
shifting some momenta, we obtain
\bea
L^{(2)}(p_1,p_2) &=& \int_{\ell_1} \, \int_{\ell_2}
\td{\ell_1} \, \td{\ell_2} \, \bigg\{  G_F(\ell_1+\ell_2+p_1) + 
G_F(\ell_1+\ell_2-p_1) + G_F(\ell_1-\ell_2-p_1) \nn \\ 
&& + \td{\ell_1+\ell_2+p_1} + \td{\ell_1+\ell_2-p_1} \bigg\}~.
\label{eq:sunrise}
\eea

For the integration of the first loop momentum $\ell_1$, we use 
the basic integrals already calculated in Ref.~\cite{Catani:2008xa}:
\bea
\int_{\ell_1} \td{\ell_1} \, G_F(\ell_1+k)
= d_\Gamma \, 
\left[  k^2 +  i 0 \right]^{-\epsilon} \,
\left[1+\theta(k^2)\, \theta(-k_0) \, 
\left(e^{i2\pi\epsilon}-1 \right)\right]~, 
\label{eq:int1loop1}
\eea
and 
\beq
\int_{\ell_1} \td{\ell_1} \, \td{\ell_1+k}
= d_\Gamma \, \left[ k^2 +  i 0 \right]^{-\epsilon}
\theta(-k^2) \, \left(e^{i2\pi\epsilon}-1 \right)~,
\label{eq:int1loop2}
\eeq
where 
\beq
d_{\Gamma} = - \frac{c_\Gamma}{2} \,  
\frac{1}{\epsilon(1-2\epsilon)} \,  
\frac{1}{\cos(\pi\epsilon)}~, \qquad
c_{\Gamma} = \frac{\Gamma(1+\epsilon)\, \Gamma^2(1-\epsilon)}
{(4\pi)^{2-\epsilon}\, \Gamma(1-2\epsilon)}~,
\eeq
which we have reexpressed in a more suitable way
in terms of $\left[ k^2 +  i 0 \right]^{-\epsilon}$. 
Applying \Eq{eq:int1loop1} and \Eq{eq:int1loop2} to \Eq{eq:sunrise},
we find 
\bea
L^{(2)}(p_1,p_2) &=& d_{\Gamma} \,
\int_{\ell_2} \td{\ell_2} \, \bigg\{
\left[ (\ell_2+p_1)^2 +  i 0 \right]^{-\epsilon}
\left( e^{i2\pi\epsilon} + 1 \right) \nn \\ &&
+ \left[ (\ell_2-p_1)^2 +  i 0 \right]^{-\epsilon}
\left[ e^{i2\pi\epsilon} - \theta((\ell_2-p_1)^2)
\theta((\ell_2-p_1)_0)
\left(e^{i2\pi\epsilon}-1 \right)
\right] \bigg\}~.
\label{eq:sunrise2}
\eea

The new phase--space integrals that we have to evaluate now for the 
integration over the second loop momentum $\ell_2$ are then 
quite similar to those already encountered at one--loop. 
The calculation is elementary, and we obtain 
\bea
&& \!\!\!\!\!\!\!\!\!\!\!\!\!\!\!\!
d_\Gamma \,
\int_{\ell_2} \td{\ell_2} \, \left[ (\ell_2+k)^2 +  i 0 \right]^{-\epsilon}
= \nn \\ &&
-\, G_2 \, \frac{\sin(\pi \epsilon) \, e^{-i 2\pi\epsilon}}
{\sin(3 \pi \epsilon)} \,
 (-k^2-i0)^{1-2\epsilon}
\left[ 1 + \theta(k^2) \theta(-k_0)  
\left( e^{i2\pi\epsilon} - 1\right) \right]~,
\label{eq:int2loop1}
\eea
and 
\bea
&& \!\!\!\!\!\!\!\!\!\!\!\!\!\!\!\!
d_\Gamma \,
\int_{\ell_2} \td{\ell_2} \, \left[ (\ell_2+k)^2 +  i 0 \right]^{-\epsilon} \, 
\theta((\ell_2+k)^2) \, \theta((\ell_2+k)_0) = \nn \\ &&
G_2 \, \frac{\sin(\pi \epsilon)}
{\sin(3 \pi \epsilon)} \, (-k^2-i0)^{1-2\epsilon}
\left[ \theta(-k^2) - \theta(k^2) \, \theta(k_0) \, e^{-i2\pi\epsilon}
\right]~,
\label{eq:int2loop2}
\eea
where
\beq
G_2=\frac{\Gamma(-1+2\epsilon)\,\Gamma(1-\epsilon)^{3}}
{(4\pi)^{4-2\epsilon}\, \Gamma(3-3\epsilon)}~.
\eeq
Applying \Eq{eq:int2loop1} and \Eq{eq:int2loop2} to \Eq{eq:sunrise2}, 
and summing up all the theta functions, we finally get
\beq
L^{(2)}(p_1,p_2) = - \, G_2 \, (-p_1^2-i 0)^{1-2\epsilon}~,
\eeq
which is the well--known result for the massless sunrise 
two--loop two--point function.

\end{document}